\documentclass{nature}

\usepackage{aas_macros}
\usepackage{amssymb}
\title{Afterglow spectrum of a $\gamma$-ray burst with the highest
known redshift z=6.295}

\author{N.~Kawai$^{1}$,
G.~Kosugi$^{2}$,
K.~Aoki$^{3}$,
T.~Yamada$^{2}$,
T.~Totani$^{4}$,
K.~Ohta$^{4}$,
M.~Iye$^{2}$,
T.~Hattori$^{3}$,
W.~Aoki$^{2}$,
H.~Furusawa$^{3}$,
K.~Hurley$^{5}$,
K.~Kawabata$^{6}$,
N.~Kobayashi$^{7}$,
Y.~Komiyama$^{3}$,
Y.~Mizumoto$^{2}$,
K.~Nomoto$^{8}$,
J.~Noumaru$^{3}$,
R.~Ogasawara$^{2}$,
R.~Sato$^{1}$,
K.~Sekiguchi$^{3}$,
Y.~Shirasaki$^{2}$,
M.~Suzuki$^{9}$,
T.~Takata$^{2}$,
T.~Tamagawa$^{9}$,
H.~Terada$^{3}$,
J.~Watanabe$^{2}$,
Y.~Yatsu$^{1}$, and
A.~Yoshida$^{10}$
}

\begin{document}

\maketitle

\begin{affiliations}
\item Department of Physics, Tokyo Institute of Technology, 
2-12-1 Ookayama, Meguro-ku, Tokyo 152-8551, Japan.

\item National Astronomical Observatory of Japan, Osawa 2-21-1,
Mitaka,  Tokyo 181-8588 Japan.

\item Subaru Telescope, National Astronomical Observatory
of Japan, Hilo, HI 96720, USA.

\item Department of Astronomy, Kyoto University, Sakyo-ku, Kyoto, 606-8502, Japan.

\item University of California at Berkeley,
Space Sciences Laboratory,  7 Gauss Way, Berkeley, CA, 94720-7450, USA.

\item Department of Physical Science, Hiroshima University, 1-3-1 Kagamiyama, Higashi-Hiroshima, Hiroshima 739-8526, Japan.

\item Institute of Astronomy, University of Tokyo, 2-21-1 Osawa, Mitaka, Tokyo 181-0015, Japan

\item Department of Astronomy, University of Tokyo, Hongo 7-3-1,
 Bunkyo-ku, Tokyo 113-0033, Japan

\item RIKEN (Institute of Physical and Chemical Research),
2-1 Hirosawa, Wako, Saitama 351-0198, Japan.

\item Department of Physics, Aoyama Gakuin University,
Sagamihara, Kanagawa 229-8558, Japan.
\end{affiliations}

\begin{abstract}
Gamma-ray bursts (GRBs) and their afterglows have been proposed as an
excellent probe to study the evolution of cosmic star
formation, 
the reionization of the intergalactic medium, and the metal enrichment
history of the universe
\cite{1997ApJ...486L..71T,1998ApJ...501...15M,2000ApJ...536....1L,
2000ApJ...540..687C}, since the prompt $\gamma$-ray emission of GRBs
should be detectable\cite{2000ApJ...536....1L} out to distances $z>10$.
Hitherto, the highest measured redshift for a GRB has been
$z=4.50$\cite{2000A&A...364L..54A}. 
Here we report the optical spectrum of the
afterglow of GRB 050904 obtained 3.4 days after the burst.  The spectrum
shows a clear continuum at the long wavelength end of the spectrum with
a sharp cutoff at around 9000 \AA\ due to Ly$\alpha$ absorption  at a
redshift of 6.3 with a damping wing.
Little flux is present in the waveband shortward of the Ly$\alpha$ break.
A system of absorption lines of heavy elements at
redshift $z=6.295\pm 0.002$ were also detected, yielding a 
precise measurement of the largest known redshift of a GRB.
Analysis of the Si~{\small II} fine structure lines suggest a
dense metal-enriched environment around the GRB progenitor, providing
unique information on the properties of the gas in a galaxy when
the universe was younger than one billion years.
\end{abstract}

GRB 050904 was a long burst (duration $T_{90}=225$ s) detected by the Swift
$\gamma$-ray burst satellite on 4 September 2005, 
01:51:44 UT\cite{2005GCN..3910....1C,2005GCN..3938....1S}.  
Its position was immediately disseminated
via the GRB Coordinates Network (GCN).  While the optical observations
at the Palomar 60$^{\prime\prime}$ telescope carried out 3.5 hours after the
trigger did not 
reveal a new source with upper limits of $R>20.8$ mag and $I>19.7$
mag\cite{2005GCN..3912....1F},         
a relatively bright near infrared source with $J\sim17.5$ mag
was detected three hours after the burst in the Swift X-ray telescope error
circle\cite{2005GCN..3913....1H},      
which showed a temporal decay with an index of $-1.20$, fully consistent with
being a typical GRB afterglow.  Analysis of the near-infrared colors,
combined with the non-detection in the optical bands, led to the suggestion
that the burst originated at a high redshift\cite{2005GCN..3914....1H}, 
$5.3<z<9.0$. 
A refined photometric redshift $z=6.10 ^{+0.37}_{-0.12}$ was
reported based on ESO VLT observations in the $J$, $H$, $K$, and $I$
bands\cite{2005GCN..3924....1A}.       

%
%
We observed the field of GRB 050904 with the Faint Object Camera And
Spectrograph (FOCAS)\cite{2002PASJ...54..819K}  
on the 8.2m Subaru Telescope atop Mauna Kea starting on the night of 6
September. 
In the $z^{\prime}$ band image (600 s exposure, mid-epoch 7 September, 8:04 UT),
we detected the afterglow at
$z^{\prime}{\rm (AB)}=23.71\pm 0.14$ mag, but we failed to detect it in the
$I_C$ band even with a longer exposure (900 s), which implied that the Lyman break should be present around
$\lambda\approx$8500--9000 \AA. 

We then obtained a grism spectrum of the afterglow candidate, 
which exhibited a sharp cutoff at $\lambda\approx$9000 \AA\  with 
strong depletion of the continuum at shorter wavelengths,
strikingly similar to the spectra of
quasars\cite{2003AJ....126....1W} 
at $z>6$ except for the absence of a broad Ly$\alpha$ emission line.
The emission is very weak in the wavelength range shorter than 8900
\AA.  In particular the flux is consistent with zero in the ranges
8500--8900 \AA\ and 7000--7500 \AA\ that extend shortward of the
Ly$\alpha$ and Ly$\beta$ wavelengths for $z \sim 6.3$.
This is a clear signature of absorption by
neutral hydrogen in the intergalactic medium at $z>6$, and marks the
first detection of a Gunn-Peterson trough\cite{1965ApJ...142.1633G} from
an object other than high-$z$ quasars\cite{2001AJ....122.2850B}.
We also find weak emission features at $\sim$7500--8300 \AA\,  
which are presumably leakage flux from the continuum emission that is also
found in quasar spectra at similar redshifts.

At the longer wavelength end of the spectrum is a flat continuum with a
series of absorption lines, which we identify as S {\small II}, Si
{\small II}, O {\small I}, and C {\small II} lines at a common redshift
of $z=6.295 \pm 0.002$. We believe that this is the redshift of the GRB host
galaxy, since no other absorption line system was observed at a redshift
consistent with that of the Ly$\alpha$ break.
This firm spectroscopic identification of the
redshift breaks the previous record of GRB 000131 at
$z$=4.50\cite{2000A&A...364L..54A}.   

From a closer examination of the absorption lines,
we find that they are not saturated and can be used to estimate the column
densities of the heavy elements as shown in Table 1.
Using the standard photospheric solar abundances\cite{1998SSRv...85..161G}
we obtain the metalicity of these elements  as [C/H]=$-2.4$, [O/H]=$-2.3$,
[Si/H]=$-2.6$, and [S/H]=$-1.0$,
where $\log N_{\rm HI} ({\rm cm}^{-2})\sim21.3$ is assumed
based on a DLA (damped Lyman $\alpha$ system) model for the Ly$\alpha$
damping wing presented below.
These values may not represent the typical abundances in the GRB host
galaxy for several reasons.  First, they are derived using only a single
ionization state for each element.  Depletion due to dust condensation
may modify the Si abundance in particular.  And second, the spatial
distribution of the heavy elements may be significantly different from
that of hydrogen.  It is possible that the heavy elements are
distributed only locally around the GRB source in a
metal-enriched circumstellar shell, while the neutral hydrogen is
distributed on a larger scale in or outside the host galaxy.

Further analysis of the Si lines allows us to constrain the scale of the
absorbing metals.  Using the equivalent width ratio of the fine structure
transition lines Si~{\small II}$\ast$ $\lambda$1264.7 and 
Si~{\small II} $\lambda$1260.4, the electron density can be
constrained\cite{2002MNRAS.329..135S} as
$\log n_e ({\rm cm}^{-3})= 2.3 \pm 0.7$ for a reasonable temperature range of
$10^3\,{\rm K} <T< 10^5\, {\rm K}$.
Combined with the column density and the abundance of Si derived above
and assuming a hydrogen ionization fraction of 0.1, we
obtain the physical depth of the absorbing system to be 0.4 pc with an
uncertainty of a factor of $\sim10$, reflecting the statistical errors
and the possible temperature range.
These fine structure lines have been found in GRB afterglow spectra
\cite{2004A&A...419..927V,2005astro.ph..8270C,2005astro.ph.11498B},
whereas they have never been
clearly detected in QSO-DLAs\cite{2004A&A...419..927V}. 
This is consistent with a local origin for the absorption
such as a metal-enriched molecular cloud in the star-forming
region or a dense metal-enriched shell nebula swept-up by a progenitor
wind prior to the GRB onset suggested for GRB
021004\cite{2003ApJ...595..935M,2003ApJ...588..387S}
and GRB 030226\cite{2004AJ....128.1942K}.  
The column density of C~{\small II} is also consistent with the
calculation for a carbon-rich Wolf-Rayet wind\cite{2005astro.ph..7659V}.

As shown in Fig 1, the Ly$\alpha$ cutoff exhibits the signature of a
damping wing redward of the Ly$\alpha$ wavelength.  This is the first
detection of significant neutral hydrogen absorption at $z\gtrsim 6$,
and offers the first case for which one can explore the distribution of
neutral hydrogen in the vicinity of a GRB, in the host galaxy, and/or in
intergalactic space at such high redshifts.  Such a study is
difficult with high-$z$ quasars due to their enormous UV flux which
ionizes the surrounding environment, and due to the presence of a strong
Ly$\alpha$ emission line.

There are two possibilities for the nature of the absorber.
It may be a DLA associated
with the host galaxy, which has been observed in the afterglows of
several GRBs at lower
redshifts\cite{2004A&A...419..927V,2005astro.ph..8270C,2005astro.ph.11498B}.
The other possibility is the neutral hydrogen in the
intergalactic medium (IGM) left over from the pre-reionization
era\cite{1998ApJ...501...15M}.  If the latter is the case, we can measure
the neutral fraction of the IGM at $z \gtrsim 6$ for the first time, giving
crucially important information on the reionization history of the
universe.  
We find that the wing shape can be reproduced either by DLA absorption 
(see inset of Fig. 1) or by the IGM.
A comprehensive spectral fitting analysis is necessary to
examine these possibilities, which is beyond the scope of this letter.

With the detection of metal absorption lines, we have 
shown that GRBs are found in metal-enriched regions even at such an
early phase of the Universe  as $z>6$.  It is, therefore,
possible that one would detect the metal absorption lines even from GRBs
originating from the metal-free first generation stars, as their
environment may be self-polluted by pre-burst winds as suggested by
the present observation.
One can expect to obtain afterglow spectra with much higher quality for
GRBs at even higher redshifts in the immediate future, considering that
Swift is constantly localizing faint GRBs, and that our spectrum was
taken when the afterglow had faded by more than an order of magnitude
since its first detection in the $J$ band.
Such future data will
give us even better opportunities to probe the formation of stars
and galaxies in the early universe.

\newpage
\begin{figure}
\includegraphics[width=1.0\textwidth]{fig1_aoki2.ps} \caption{
Spectrum of the afterglow of GRB 050904 covering wavelengths 7000--10000
\AA\ with a resolution $R=\lambda / \Delta\lambda\sim1000$ at 9000 \AA.
It was taken with Subaru/FOCAS at mid-epoch 7 September, 12:05 UT, 3.4
days after the burst for a total exposure of 4.0 hours.  The abscissa is
the observed wavelength converted to that in vacuum.  The spectrum is
smoothed to a resolution $R=\lambda / \Delta\lambda\sim 600$ at 9000
\AA.  The locations of the identified absorption lines (see Table 1) at
$z=4.840$ and $z=6.295$, as well as the wavelengths of Ly$\alpha$ and
Ly$\beta$ at $z=6.295$ are shown with vertical dotted lines.  The
one-sigma errors are shown with an offset of -1.0 at the bottom of the
panel.  
In the inset, we show a model for the damping wing of
Ly$\alpha$ absorption with a neutral hydrogen column density 
$\log N_{\rm HI} ({\rm cm}^{-2})=21.3$ at a redshift $z=6.3$
by the solid line, overlaid on the observed spectrum in the wavelenghth
range of 8700--9500  \AA.
The dotted line indicates the unabsorbed continuum model following a
power-law ($f_\nu \propto \nu^{-1}$) function
as typically observed for GRB afterglows.
Note that only the red wing is relevant to the
fit, since the emission blueward of Ly$\alpha$ is absorbed by the IGM.
}
\end{figure}


\bibliography{Subaru050904}


\begin{addendum}
\item Based on data collected at the Subaru Telescope, which is operated
   by the National Astronomical Observatory of Japan.  We are grateful for the
   excellent support by the observatory.  N.K. acknowledges the support
   by Grants-in-Aid for Scientific Research from the Ministry of
   Education, Culture, Sports, Science and Technology of Japan and the
   Tokyo Tech COE-21 program ``Nanometer-scale Quantum Physics''.  We
   thank S. Barthelmy for maintaining the GCN, and the Swift team for
   providing rapid GRB localizations.
 \item[Competing Interests] The authors declare that they have no
competing financial interests.
 \item[Correspondence] Correspondence and requests for materials
should be addressed to \\
N.K. (email: nkawai@phys.titech.ac.jp).
\end{addendum}

\begin{table}
\begin{center}
\begin{tabular}{ccccc}
\hline
 Observed         &  Equivalent     &   Column      &  Line                        & Redshift\\[-12pt]
 Wavelength       &  Width          &   Density     &  Identification              & \\[-12pt]
  (\AA)           &  (\AA)          & log (cm$^{-2}$) &                              & \\
\hline
9041.0 $\pm$ 0.8  &  4.5 $\pm$ 1.0  & 14.44$^{\sf +0.14}_{\sf -0.16}$ &  C {\small IV} $\lambda$1548.2       & 4.840$\pm$0.001 \\[-12pt]
                  &                 &           --          --        & (N {\small V} $\lambda$1238.8)       &(6.298$\pm$0.001) \\
9055.9 $\pm$ 1.7  &  1.7 $\pm$ 1.0  & 14.21$^{\sf +0.25}_{\sf -0.46}$ &  C {\small IV} $\lambda$1550.8       & 4.840$\pm$0.001 \\[-12pt]
                  &                 &           --          --        & (N {\small V} $\lambda$1242.8)       &(6.287$\pm$0.001)\\
9146.4 $\pm$ 1.8  &  3.8 $\pm$ 1.1  & 15.60$^{\sf +0.14}_{\sf -0.17}$ &  S {\small II} $\lambda$1253.8        & 6.295$\pm$0.001 \\
9188.7 $\pm$ 2.6  &  6.1 $\pm$ 3.7  & 16.20$^{\sf +1.87}_{\sf -0.92}$ &  S {\small II} $\lambda$1259.5       & 6.295$\pm$0.002 \\
9195.9 $\pm$ 1.2  &  8.3 $\pm$ 2.7  & 14.29$^{\sf +0.57}_{\sf -0.39}$ & Si {\small II} $\lambda$1260.4       & 6.296$\pm$0.001 \\
9225.8 $\pm$ 1.8  &  3.9 $\pm$ 1.1  & 13.63$^{\sf +0.13}_{\sf -0.16}$ & Si {\small II}$^{*}$ $\lambda$1264.7 & 6.295$\pm$0.001 \\
9499.1 $\pm$ 0.9  & 10.3 $\pm$ 1.9  & 15.85$^{\sf +0.39}_{\sf -0.28}$ &  O {\small I} $\lambda$1302.2        & 6.295$\pm$0.001 \\
9737.2 $\pm$ 1.1  & 12.3 $\pm$ 2.4  & 15.41$^{\sf +0.30}_{\sf -0.26}$ &  C {\small II} $\lambda$1334.5       & 6.296$\pm$0.001 \\
\hline
\end{tabular}
\end{center}
\caption{List of the absorption lines detected in the spectrum of the
 optical afterglow of GRB 050904. The wavelengths and equivalent widths
 were derived by fitting a single Gaussian.
 The column densities of lines were estimated by the standard curve of
 growth analysis\cite{1978ppim.book.....S}. 
 The equivalent widths in the table are observed ones, i.e., not converted to
 the rest-frame.
 The quoted uncertainties are 1$\sigma$ statistical errors.
 Most of the absorption lines are consistent with being at a single
 redshift of $z=6.295 \pm 0.002$ within the statistical uncertainties.
 The absorption lines at $\lambda=$9041.0 and 9055.9 \AA\ could be
 identified as N~{\small V} $\lambda\lambda$1238.8,1242.8 if they
 are at a similar redshift as the other absorption lines.  However, the
 derived redshifts of these two lines are significantly inconsistent
 with each other.  Another possible identification is C~{\small IV}
 $\lambda\lambda$1548.2, 1550.8 in an intervening system at
 $z=4.840$, which we think is more likely.  }
\end{table}

\end{document}